\documentclass[journal=jctcce,manuscript=article,layout=twocolumn]{achemso}
\setkeys{acs}{etalmode=truncate,maxauthors=10}
\setkeys{acs}{articletitle=true}

\usepackage{chemformula} 
\usepackage[T1]{fontenc} 
\usepackage{amsmath,amsfonts}
\usepackage{physics}
\usepackage{comment}
\usepackage{booktabs}
\usepackage{multirow}
\usepackage{textcomp}  
\mciteErrorOnUnknownfalse
\usepackage{caption}
\usepackage{subcaption}
\usepackage{widetext}
\usepackage[version=4]{mhchem}
\usepackage{qcircuit}
\usepackage{enumitem}

\mathchardef\mhyphen="2D

\newcommand{\HF}{\mathrm{HF}}
\newcommand{\CIS}{\mathrm{CIS}}
\newcommand{\CNOT}{\mathrm{CNOT}}

\author{Joel Bierman}
\affiliation[Duke]{Department of Physics, Duke University}
\author{Yingzhou Li}
\affiliation[Fudan]{School of Mathematical Sciences, Fudan University}
\email{yingzhouli@fudan.edu.cn}
\author{Jianfeng Lu}
\email{jianfeng@math.duke.edu}
\affiliation[Duke]{Department of Mathematics, Duke University}
\alsoaffiliation{Department of Physics, Duke University}
\alsoaffiliation{Department of Chemistry, Duke University}

\title[Orbital Optimization]{Qubit Count Reduction by Orthogonally-Constrained Orbital Optimization for Variational Quantum
Excited States Solvers}

\abbreviations{}

\begin{document}

\begin{abstract}
We propose a state-averaged orbital optimization scheme for improving the
accuracy of excited states of the electronic structure Hamiltonian for use
on near-term quantum computers. Instead of parameterizing the orbital rotation operator in the conventional fashion as an exponential of an anti-hermitian matrix, we parameterize the orbital rotation as a general partial unitary matrix. Whereas conventional orbital optimization methods minimize the state-averaged energy using successive Newton steps of the second-order Taylor expansion of the energy, the method presented here optimizes the state-averaged energy using an orthogonally-constrained gradient projection method which does not require any expansion approximations. Through extensive benchmarking of the
method on various small molecular systems, we find that the method is
capable of producing more accurate results than fixed basis FCI while
simultaneously using fewer qubits. In particular, we show that for
\ch{H2},  the method is capable of matching the accuracy of FCI in the
cc-pVTZ basis (56 qubits) while only using 14 qubits.
\end{abstract}

\section{Introduction}\label{sec:intro}

One of the early applications for quantum computers is expected to be the
electronic structure problem~\cite{helgaker2014molecular}, however, the
error stemming from the basis set truncation in the second quantization
formulation will likely present a major obstacle for realizing accurate
solutions for academically and industrially relevant chemical
systems~\cite{kuhnAccuracyResourceEstimations2019,
elfvingHowWillQuantum2020, gonthierMeasurementsRoadblockNearterm2022}. If
no resource reduction techniques are employed, one qubit is needed for
each spin-orbital. As a result of the limited number of qubits on current
hardware, experimental demonstrations have been limited to small molecules
represented by small basis
sets~\cite{kandalaHardwareefficientVariationalQuantum2017,
shenQuantumImplementationUnitary2017,
omalleyScalableQuantumSimulation2016,
ganzhornGateEfficientSimulationMolecular2019}. Several methods have been
developed for more compact basis set representations in both the classical
and quantum settings. Explicitly correlated
methods~\cite{mcardleImprovingAccuracyQuantum2020,
sokolovOrdersMagnitudeReduction2023, mottaQuantumSimulationElectronic2020,
kumarQuantumSimulationMolecular2022} apply a similiarity transformation to
the problem Hamiltonian that has an explicit dependence on the coordinates
of the electrons. The intuition behind this is that such basis sets may be
able to efficiently capture effects from electron-electron interactions
which are often the cause of the inefficiency of fixed single-particle
basis set representations~\cite{kongExplicitlyCorrelatedR122012,
kutzelniggRatesConvergencePartial1992}. Downfolded effective Hamiltonian
techniques~\cite{metcalfResourceEfficientChemistryQuantum2020,
huangLeveragingSmallScale2022, claudinoImprovingAccuracyEfficiency2021,
baumanDownfoldingManybodyHamiltonians2019,
baumanQuantumSimulationsExcited2019} take a full orbital space
Hamiltonian, unitarily transform it according to an excitation operator
that includes excitations outside of a given active space, then project it
onto the active space. Through this process, an effective Hamiltonian is
produced which includes correlation effects from outside the active space,
but which is low-dimensional and acts only on the active space. Orbital
optimization methods~\cite{biermanImprovingAccuracyVariational2023,
omiyaAnalyticalEnergyGradient2022, degraciatrivinoCompleteActiveSpace2023,
mizukamiOrbitalOptimizedUnitary2020, liOptimalOrbitalSelection2020,
tillyReducedDensityMatrix2021,
yalouzStateaveragedOrbitaloptimizedHybrid2021} introduce the elements of a
similarity transformation to be optimized in conjunction with the
parameters of an eigensolver minimization problem. These two minimization
problems are typically solved in an alternating fashion until some
stopping criteria are reached. Orbital optimization schemes such as
quantum CASSCF~\cite{tillyReducedDensityMatrix2021} operate by representing the orbital rotation as an exponential of an $M \times M$ anti-hermitian matrix, approximating this exponential operator by a second-order Taylor expansion, performing successive Newton steps using this approximation until convergence of the energy is achieved, then choosing an active space of $N < M$ orbitals in which to solve for the eigenvalues and eigenstates. In
this work we extend the
OptOrbVQE~\cite{biermanImprovingAccuracyVariational2023} method previously proposed by the authors, which finds the
ground state in an optimized basis, to the problem of finding excited
states of electronic structure Hamiltonians. The main differences between
OptOrbVQE and other orbital optimization ground state solvers can be summarized as follows. Instead of parameterizing the orbital rotation operator as an $M \times M$ exponential operator and carrying out successive Newton steps, we parameterize it directly as an $M \times N$ partial unitary matrix. This allows us to take advantage of modern optimization techniques that have been developed in recent years which have orthogonality constraints built in.~\cite{wenFeasibleMethodOptimization2013, gaoParallelizableAlgorithmsOptimization2019,zhangGradientTypeOptimization2014, huangBroydenClassQuasiNewton2015,gaoNewFirstOrderAlgorithmic2018a} Such optimization methods render the conventional exponential parameterization as one choice of parameterization rather than a strict requirement. We note that aside from access to a wider range of optimization techniques, this approach comes with other potential advantages. The first is that the partial unitary transformation has $NM$ constrained parameters, rather than the $(M-1)(M-2)/2$ independent parameters in the anti-hermitian matrix that appears in CASSCF. This results in a net reduction of the size of the parameter space involved for many problems, particularly when $M \gg N$. This may ease the convergence procedure. Additionally, the partial unitary nature of the parameterization implies that unlike CASSCF, we do not need to resort to the use of heuristics or chemical intuition when choosing the post-optimization active space. The dimensionality of the partial unitary matrix handles this automatically during the optimization procedure. We also note that there is a sense in which the dimensionality of this partial unitary operator is a degree of freedom which we can control. For example, we still retain the option to optimize a full $M \times M$ unitary matrix using orthogonally-constrained optimizers and use heuristics to choose the active space of $N$ orbitals. Alternatively, we could optimize over the set of $M \times m$ partial unitaries for $N < m < M$ and use heuristics to choose an active space of $N$ orbitals from the space of $m$ orbitals. This flexibility can be seen as an advantage in and of itself.  Furthermore, the use of an orthogonally-constrained projected gradient descent method~\cite{gaoNewFirstOrderAlgorithmic2018a} in conjunction with this orbital optimization procedure has been numerically demonstrated to be more adept at avoiding local minima and achieving better accuracy than CASSCF in a previous work by two of the authors of this work in a classical computing context.~\cite{liOptimalOrbitalSelection2020} This offers a clear motivation for the continued study of this method in a broader range of contexts such as generalization to excited states and incorporation into quantum eigensolvers. For example, in a previous work by the authors, it was demonstrated that the convergence quality of state-averaged eigensolvers such as SSVQE in a fixed basis is highly sensitive to the ansatz expressiveness and choice of circuit initialization (compared to the ground state VQE problem and the overlap-based qOMM excited states solver).\cite{biermanQuantumOrbitalMinimization2022} In this work we investigate the extent to which analogous observations hold true for the orbital-optimized case as well. Given that there is potentially a non-trivial interplay between the choice of basis (which is furthermore variable in orbital optimization) and the ability of any given ansatz to express the solution, this relation should be investigated.

\section{Excited States Quantum Eigensolvers}
\label{sec: Excited States Solvers}

Hybrid quantum-classical variational methods for finding eigenvalues of
chemical Hamiltonians operate by classically minimizing an objective
function constructed from quantities measured on a quantum computer. For
example, to find the ground state of a Hamiltonian $\hat{H}$ we would
first prepare a parametrized state $\ket{\psi(\pmb{\theta})}$ on the
quantum computer, measure the expectation value of $\hat{H}$, and carry
out the minimization problem:
\begin{equation}
    \min_{\pmb{\theta}} \bra{\psi(\pmb{\theta})} \hat{H}
    \ket{\psi(\pmb{\theta})}
\end{equation}
classically. This is the original formulation of the variational quantum
eigensolver~\cite{peruzzoVariationalEigenvalueSolver2014,
tillyVariationalQuantumEigensolver2022} (VQE). In order to extend this
method to low-lying excited states, the mutual orthogonality of these
states must be accounted for. Several methods have been proposed that
accomplish this.
SSVQE~\cite{nakanishiSubspacesearchVariationalQuantum2019} and
MCVQE~\cite{parrishQuantumComputationElectronic2019} are state-averaged
approaches which apply a parameterized circuit $\hat{U}(\pmb{\theta})$ to
a set of mutually orthogonal initial states $\{\ket{\psi_i}\}$, then
minimize an objective function of the form:
\begin{equation}\label{eq: SA-VQE objective function}
    f(\pmb{\theta}) = \sum_i w_i\bra{\psi_i}\hat{U}^{\dagger}
    (\pmb{\theta})\hat{H}\hat{U}(\pmb{\theta})\ket{\psi_i}
\end{equation}
where $\{w_i\}$ is a set of positive, real-valued weights. The main
difference between MCVQE and SSVQE is that MCVQE chooses the weights
$\{w_i\}$ to be equal, whereas SSVQE chooses them to not be equal. At
first glance this difference seems trivial, however it should be noted
that unequal weights corresponds to a global minimum comprised of the
low-lying eigenvectors, whereas an equal weighting corresponds to a global
minimum comprised of states which span the low-lying eigenspace. MCVQE
adds a classical post-processing step which diagonalizes these states in
this low-dimensional eigenspace to acquire the low-lying eigenvectors. It
is unclear which of these approaches is advantageous or if their
convergence is equivalent in practice. Other excited states methods such
as qOMM~\cite{biermanQuantumOrbitalMinimization2022} and
VQD~\cite{higgottVariationalQuantumComputation2019} take overlap-based
approaches to enforcing the mutual orthogonality of the solution by
including penalty terms in the objective function which vanish when pairs
of states are orthogonal. Thus, the orthogonality is enforced only at the
global minimum rather than at every point in the cost function landscape.

\section{State-Averaged Orbital Optimization}
\label{sec: orbital optimization}

In OptOrbVQE we take the electronic structure Hamiltonian in its fermionic
second-quantization representation:
\begin{equation} \label{eq: fermionic hamiltonian}
    \hat{H} = \sum _{p,q = 1}^{M} h_{pq}\hat{a}^{\dagger}_{p}\hat{a}_q
    + \frac{1}{2}\sum_{p,q,r,s = 1}^{M} v_{pqrs}
    \hat{a}^{\dagger}_{p}\hat{a}^{\dagger}_{q}\hat{a}_{s}\hat{a}_{r},
\end{equation}
and rotate the set of $M$ orbitals $\{\psi_1, \psi_2, ..., \psi_M\}$ according to the
partial unitary transformation $\hat{V}$:
\begin{equation} \label{eq: basis func transform}
    \Tilde{\psi_{i}} = \sum_{j}^{M} \hat{V}_{ji}\psi_{j}
\end{equation}
resulting in a new set of $N < M$ orbitals $\{\tilde{\psi}_1,
\tilde{\psi}_2, \dots, \tilde{\psi}_N\}$. This is equivalent to
transforming the Hamiltonian as:
\begin{equation}
    \begin{split}
        &\Tilde{H}(\hat{V}) = \sum_{p^{\prime}, q^{\prime} = 1}^{N}
        \sum_{p,q = 1}^{M} h_{pq}\hat{V}_{pp^{\prime}}\hat{V}_{qq^{\prime}}
		\tilde{a}^{\dagger}_{p^{\prime}}\tilde{a}_{q^{\prime}}\\
        &+ \frac{1}{2}\sum_{p^{\prime},q^{\prime},r^{\prime},
		s^{\prime}=1}^{N}\sum_{p,q,r,s=1}^{M}v_{pqrs}\hat{V}_{pp^{\prime}}
		\hat{V}_{qq^{\prime}}\hat{V}_{ss^{\prime}}\hat{V}_{rr^{\prime}}
		\tilde{a}^{\dagger}_{p^{\prime}}\tilde{a}^{\dagger}_{q^{\prime}}
		\tilde{a}_{s^{\prime}}\tilde{a}_{r^{\prime}.}
    \end{split}
\end{equation}
The orbital optimization then corresponds
to minimizing the expectation value of this Hamiltonian with respect to a
fixed quantum state $\hat{U}(\pmb{\theta})\ket{\psi_{ref}}$ provided by a
quantum eigensolver. The total minimization problem is then given by:
\begin{equation} \label{eq: total gs minimization problem}
    \min_{\substack{\pmb{\theta}\\
    \hat{V}\in\mathcal{U}(M,N)}} \bra{\psi_{\text{ref}}}
	\hat{U}^{\dagger}(\pmb{\theta})\Tilde{H}(\hat{V})
	\hat{U}(\pmb{\theta})\ket{\psi_{\text{ref}}}
\end{equation}
where $\mathcal{U}(M,N)$ is the set of $M \times N$ real partial unitary
matrices. The simplest way to generalize this problem is to consider
Eq.~\ref{eq: SA-VQE objective function} to be a function of both
$\pmb{\theta}$ and $\hat{V}$:
\begin{equation}\label{eq: SA optorb objective function}
    f(\pmb{\theta}, \hat{V}) = \sum_{i} w_i\bra{\psi_{\text{ref},i}}
	\hat{U}^{\dagger}(\pmb{\theta})\Tilde{H}(\hat{V})\hat{U}(\pmb{\theta})
	\ket{\psi_{\text{ref},i}}
\end{equation}
and minimize the resulting state-averaged analog problem of Eq.~\ref{eq:
total gs minimization problem}:
\begin{equation}\label{eq: total es minimization problem}
    \min_{\substack{\pmb{\theta}\\
    \hat{V}\in\mathcal{U}(M,N)}} f(\pmb{\theta}, \hat{V}).
\end{equation}
Such state-averaged analogs of CASSCF-like orbital optimization schemes~\cite{yalouzStateaveragedOrbitaloptimizedHybrid2021,
omiyaAnalyticalEnergyGradient2022} have previously been explored in the
literature. Thus, we expect a state-averaged analog of OptOrbVQE to also
perform well. It is worth noting that an overlap-based orbital
optimization objective function has been proposed in the classical
literature~\cite{yalouzOrthogonallyConstrainedOrbital2023}, which allows
for a separate optimal basis to be computed for each excited state. The
authors claim that this allows for more accurate excitation energies to be
computed. The method assumes the availability of the CI coefficients found by the
eigensolver, which would require exponentially-expensive full state
tomography to acquire in the quantum computing setting.

The total minimization problem  Eq.~\ref{eq: total es minimization
problem} is divided into two subproblems: minimization with respect to the
ansatz parameters $\pmb{\theta}$ and minimization with respect to
$\hat{V}$. These two subproblems are solved in an alternating fashion,
where one is fixed while the other is varied. The optimal parameters for
one subproblem are then used for the initialization of the next run of the
other until some global stopping criteria are met. For example, for a
given optimal $\hat{V}$ we can compute $\tilde{H}(\hat{V})$ and carry out
a quantum excited states solver to find an optimal $\pmb{\theta}$ in the
rotated basis. For a given $\pmb{\theta}$ found by a quantum excited
states solver, we can compute the 1 and 2-RDMs with respect to each state
in the set of computed excited states
$\{\hat{U}(\pmb{\theta})\ket{\psi_{\text{ref},i}}\}$, then vary
Eq.~\ref{eq: total es minimization problem} with respect to $\hat{V}$. The optimization with respect to $\pmb{\theta}$ is handled via one of several known quantum excited states solvers such as SSVQE~\cite{nakanishiSubspacesearchVariationalQuantum2019}, MCVQE~\cite{parrishQuantumComputationElectronic2019}, VQD~\cite{higgottVariationalQuantumComputation2019}, or qOMM~\cite{biermanQuantumOrbitalMinimization2022}. The optimization with respect to $\hat{V}$ (keeping $\pmb{\theta}$ fixed) is carried out using an orthogonally-constrained optimization procedure. In this work we use an orthogonally-constrained projected gradient method~\cite{gaoNewFirstOrderAlgorithmic2018a}, which has a parameter update step defined as:

\begin{equation}
    \hat{V}_{n+1} = \mathrm{orth}(\hat{V}_n - \eta \nabla f(\hat{V}_n))
\end{equation}where $\nabla f(\hat{V}_n))$ is the gradient of Eq.~\ref{eq: SA optorb objective function} with respect to $\hat{V}$ with fixed $\pmb{\theta}$, $\eta$ is a step size which is chosen adaptively in an alternating Barzilai-Borwein fashion, and the $\mathrm{orth}$ function is defined as:~\cite{gaoNewFirstOrderAlgorithmic2018a, liOptimalOrbitalSelection2020}

\begin{equation}
    \mathrm{orth}(A) = AQ\Lambda^{-\frac{1}{2}}Q^{\dagger}.
\end{equation}Here $Q$ is a matrix whose columns are the eigenvectors of $A^{\dagger}A$ and $\Lambda$ is a diagonal matrix whose entries are the eigenvalues of $A^{\dagger}A$.
As
was done for OptOrbVQE, we explicitly state the super and subscript
notation used for the total problem to avoid confusion:

\begin{itemize}
    \item The subscript $l$ will index the iteration number in the
    minimization problem where $\hat{V}$ is varied.

    \item The subscript $m$ will index the iteration number in the
    minimization problem where $\pmb{\theta}$ is varied.

    \item The subscript $n$ will index a global ``outer loop'' iteration
    number that characterizes how many times both subproblems have been
    carried out.

    \item The superscript $\emph{opt}$ will denote the optimal parameter
    found in each subproblem for a given outer loop iteration number.
\end{itemize}
We now give an explicit step-by-step procedure for the total problem:
\begin{enumerate}
    \item Set $n = 0$. Choose an initial partial unitary
    $\hat{V}_{n=0,l=0}$, an initial set of ansatz parameters
    $\pmb{\theta}_{n=0,m=0}$, and a stopping threshold $\epsilon$.

    \item Calculate $\tilde{H}(\hat{V})$ on a classical computer and run a
    quantum eigensolver algorithm to obtain $\pmb{\theta}_{n}^{opt}$.

    \item If $\lvert f(\pmb{\theta}_{n}^{opt}, \hat{V}_{n-1}^{opt}) -
    f(\pmb{\theta}_{n-1}^{opt}, \hat{V}_{n-2}^{opt})\rvert < \epsilon$,
    halt the algorithm. Else, continue to the next step.

    \item Measure the 1 and 2-RDMs with respect to the set of states
    $\{\hat{U}(\pmb{\theta}_{n}^{opt})\ket{\psi_{\text{ref},i}}\}$ on a
    quantum computer.

    \item Using the 1 and 2-RDMs from the previous step, minimize
    Eq.~\ref{eq: total es minimization problem} with respect to
    $\hat{V}_{n}$ to obtain $\hat{V}_{n}^{opt}$.

    \item Set $n = n + 1$, $\hat{V}_{n+1,l=0} = \hat{V}_{n}^{opt}$, and
    $\pmb{\theta}_{n+1,m=0} = \pmb{\theta}_{n}^{opt}$. Optionally, a small
    random perturbation can be added to the latter two quantities. Repeat
    steps 2-6.
\end{enumerate}
Step 5 requires the use of a classical optimizer which constrains
$\hat{V}$ to be a partial unitary. Several methods which do this
exist~\cite{wenFeasibleMethodOptimization2013,
gaoParallelizableAlgorithmsOptimization2019,
zhangGradientTypeOptimization2014, huangBroydenClassQuasiNewton2015}, but
in our work we use an orthogonal projection
method~\cite{gaoNewFirstOrderAlgorithmic2018a}. In general, $\pmb{\theta}$
and $\hat{V}_{n+1,l=0}$ could be any real vector and real partial unitary,
respectively, however it is intuitive to use information from the $n$th
outer loop iteration to inform this choice. In our work we choose
$\pmb{\theta}_{n+1,l=0} = \pmb{\theta}_{n}^{opt}$ and $\hat{V}_{n+1,l=0} =
\mathrm{orth}(\hat{V}_n^{opt} + \mathrm{Rand(M,N))}$, where $\mathrm{Rand(M,N)}$ is an $M \times N$ matrix whose elements are
sampled from a normal distribution with average 0 and standard deviation
0.01. Additionally, although Eq.~\ref{eq: total es minimization problem}
is written as a state-averaged function of $\pmb{\theta}$, step 2 does not
necessarily need to be carried out using a state-average quantum
eigensolver. The only requirement is that the solver returns solution
states to be used for the calculation of 1 and 2-RDMs. Overlap-based
methods such as qOMM~\cite{biermanQuantumOrbitalMinimization2022} and
VQD~\cite{higgottVariationalQuantumComputation2019} could be used, however
for our work we test MCVQE~\cite{parrishQuantumComputationElectronic2019}
and SSVQE~\cite{nakanishiSubspacesearchVariationalQuantum2019}. We further note that because the method derives an optimized basis of $N < M$ orbitals from an initial large basis of $M$ orbitals, it cannot match or exceed the accuracy obtained by solving the eigenvalue problem in the full $M$ orbital active space. This is intuitive from the perspective that the ansatz and orbital rotation parameters constitute a joint variational space. Restricting the variational space will in general restrict the maximum attainable accuracy. In particular, it is known that orbital optimization methods in general are not efficient at capturing correlation effects arising from Coulomb repulsion between electrons. For this, one could consider combining the orbital optimization scheme with other methods such as explicitly correlated methods.~\cite{mcardleImprovingAccuracyQuantum2020,sokolovOrdersMagnitudeReduction2023,mottaQuantumSimulationElectronic2020}

\section{Numerical Results}
\label{sec: Numerical Results} 

The code used for our numerical simulations is an extension of the
functionality provided by the open source package
Qiskit~\cite{qiskitcontributorsQiskitOpensourceFramework2023a}. Qiskit
provides an implementation of VQE~\cite{
peruzzoVariationalEigenvalueSolver2014,
tillyVariationalQuantumEigensolver2022}, which we have modified to produce
implementations of
SSVQE~\cite{nakanishiSubspacesearchVariationalQuantum2019} and
MCVQE~\cite{parrishQuantumComputationElectronic2019}. The code for the
state-averaged orbital optimization is a modification of the code used in
our ground state orbital optimization work
\cite{biermanImprovingAccuracyVariational2023}, with the main modification
being the objective function to be minimized. The Qiskit package versions
used are Qiskit-Aer 0.12.0, Qiskit-Nature 0.4.5, and Qiskit-Terra 0.23.2.
The 1 and 2-body integrals are obtained through the
PySCF~\cite{sunPySCFPythonbasedSimulations2018} electronic structure
driver in Qiskit, which uses PySCF to perform a restricted Hartree-Fock
problem to obtain the un-optimized molecular integrals. Configuration
interaction circuits are obtained in two steps. First, the truncated
Hamiltonians are constructed from the 1 and 2-body integrals using the
Slater-Condon rules~\cite{helgaker2014molecular}, which are then exactly
diagonalized using NumPy~\cite{harrisArrayProgrammingNumPy2020}. The
reasons why we do not use PySCF's configuration interaction implementation
are two-fold: (1) PySCF does not have an CIS implementation and (2) We
have found that PySCF's CISD implementation does not always produce
orthogonal CI wavefunctions, with fidelity between two states being as
large as on the order of $10^{-1}$, even in the case where the
corresponding eigenvalues are not degenerate. This is problematic for
quantum algorithms such as SSVQE and MCVQE which require that the initial
states be mutually orthogonal.

This statevector can then be used to initialize a circuit using Qiskit's
arbitrary statevector initialization implementation. We note that although
this particular implementation requires the storage of an exponentially
large statevector in classical memory, in principle configuration
interaction state preparation on a quantum computer could be done in a
completely sparse manner with resources scaling polynomially with the
number of qubits. For example, it has been shown that Givens rotations are universal
for preparing chemically-motivated states with the Jordan-Wigner
mapping~\cite{arrazolaUniversalQuantumCircuits2022}. The authors also give
a general procedure for preparing an arbitrary statevector. In
Appendix~\ref{CIS state preparation} we give an explicit example of how
the particular case of arbitrary CIS statevectors can be prepared on a
quantum computer, which may be of independent interest. Whether or not an
efficient analogous procedure can be developed for CISD states is not
discussed here, however in our simulations we include CISD initializations
to investigate whether or not doing so would lead to further improvement. We also utilize an "excited Hartree-Fock" initialization that consists of the Hartree-Fock state and the lowest energy singly-excited states from it. In this paper, we will refer to this initialization as just "Hartree-Fock" or "HF". In Qiskit, one can use any ansatz circuit as a base pattern to be repeated
$n$ times, increasing the circuit depth and number of parameters by a
factor of $n$. In our simulations we use Qiskit's implementation of the
UCCSD ansatz~\cite{romeroStrategiesQuantumComputing2018} as a circuit
block pattern to be repeated for various values of $n$. We denote this as
$n$-UCCSD. The classical optimizer used for all test instances is
L-BFGS-B~\cite{byrdLimitedMemoryAlgorithm1995}, The FCI reference values
are calculated using CDFCI,~\cite{wangCoordinateDescentFull2019a}, and all orbital-optimized tests are run using Qiskit's \emph{AerSimulator} in noiseless statevector mode unless stated otherwise. In Sec.~\ref{sec: H4 binding curve} we make slight adjustments to this methodology when calculating a potential energy surface of \ch{H4}.

\subsection{\ch{H2}}
\label{sec: H2}

\begin{figure*}[htb!]
    
    \centering
    \includegraphics[width=\linewidth]{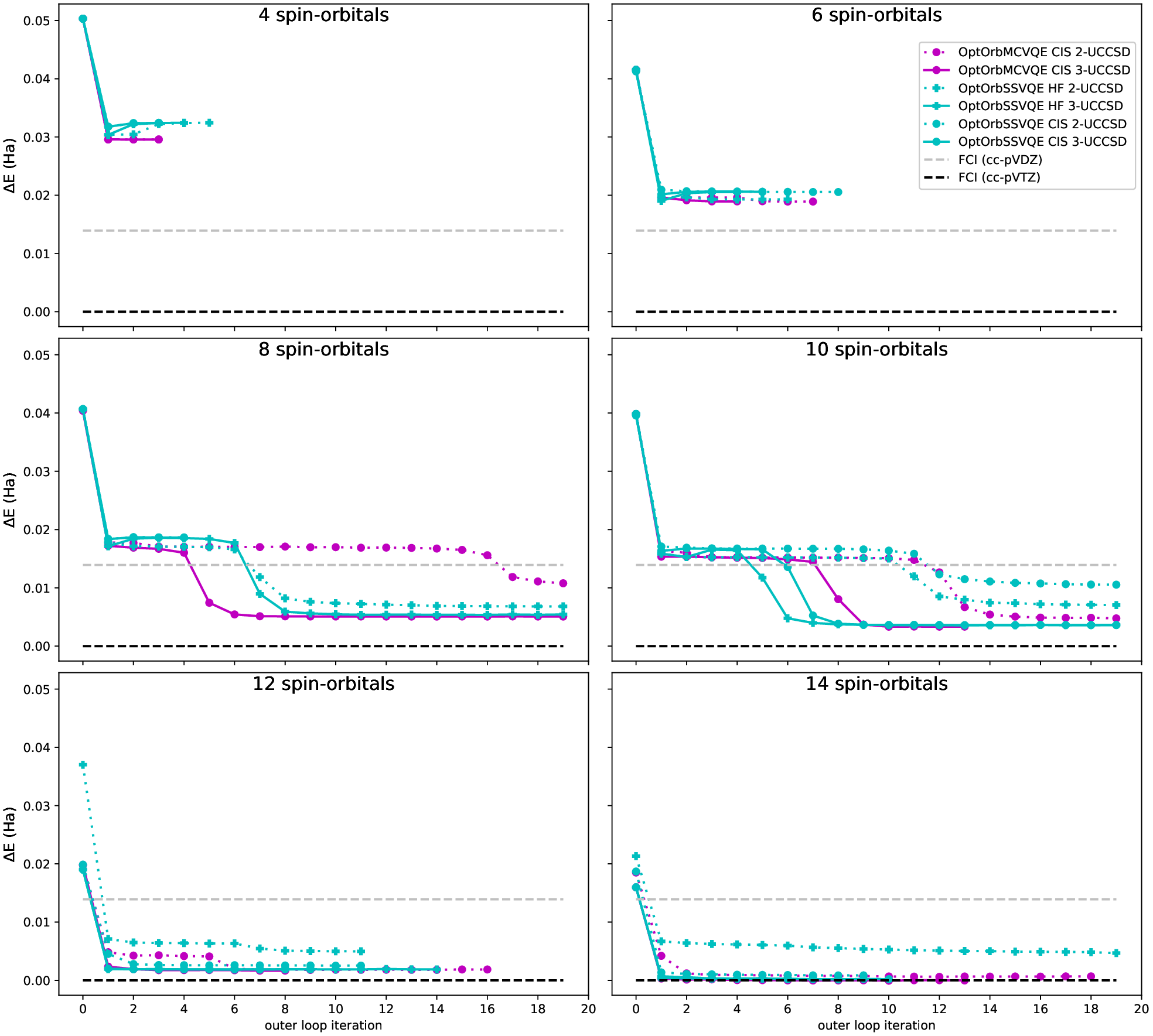}
    \caption{Convergence of orbital optimization methods for $\ch{H_2}$
    using various numbers of spin-orbitals (taken from the cc-pVQZ basis)
    as a function of the outer loop iteration. $\Delta E$ is the
    difference between the average energy and that of FCI in the cc-pVTZ
    basis (56 spin-orbitals).}
    \label{fig:H2 results}
\end{figure*}

We begin with our results for the simplest model tested, the first three energy levels of \ch{H2} at the
near-equilibrium bond distance 0.735~\AA, which are given in
Fig.~\ref{fig:H2 results}. We use cc-pVQZ (120 spin-orbitals) as the
starting basis and reduce the active space for even numbers of
spin-orbitals from 4 to 14 using the proposed orbital optimization scheme.
The difference between the average orbital optimized energy and that of
FCI (over the ground and excited states) in the cc-pVTZ basis is plotted
as a function of the outer loop iteration. Tests using both 2 and 3-UCCSD
are included to investigate the effect of increasing the ansatz
expressiveness in the algorithm. Both eigensolvers are initialized with
configuration interaction singles (CIS) states. SSVQE is additionally tested using the Hartree-Fock initialization.

It is evident that orbital optimization has the potential to achieve more
accurate average energies than FCI in the cc-pVDZ basis (20 spin-orbitals)
and can even approach cc-pVTZ quality values, but this highly depends on
the choice of eigensolver, ansatz, and number of optimized spin-orbitals.
A minimum of 8 spin-orbitals are needed to achieve a higher accuracy than
cc-pVDZ. At this point, OptOrbMCVQE can do this for both 2 and 3-UCCSD,
although OptOrbSSVQE cannot. Using 10 spin-orbitals, both eigensolvers
surpass cc-pVDZ for both 2 and 3-UCCSD, although MCVQE offers roughly a 5
milli-Hartree improvement over SSVQE for 2-UCCSD. When 3-UCCSD is used,
MCVQE offers a measurable but negligible improvement over SSVQE. At 14
spin-orbitals, cc-pVTZ quality results are achievable. Also notable is the
effect that increasing the active space has on not only the quality of
convergence, but its rate of convergence. Note that for 8 and 10
spin-orbitals, the convergence appears to plateau, hovering just above
cc-pVDZ accuracy for several iterations before rapidly surpassing it. This
behavior is not present at 12 and 14 spin-obitals, with the energy quickly
converging to near or at cc-pVTZ accuracy for the majority of tests run. As a sidenote, we note that the 14 spin-orbital tests using 3-UCCSD with a CIS initialization were stopped manually at iterations 10 and 13 for SSVQE and MCVQE, respectively as the runtime for these simulations proved to be the longest among these tests. However, we note that given that nearly all of the energy convergence occurred within the first 2 or 3 iterations, allowing the simulations to continue would likely not have resulted in further improvement.

\subsection{\ch{H4}}
\label{sec: H4}

\begin{figure}[htb]
    \centering
    \includegraphics[width=\linewidth]{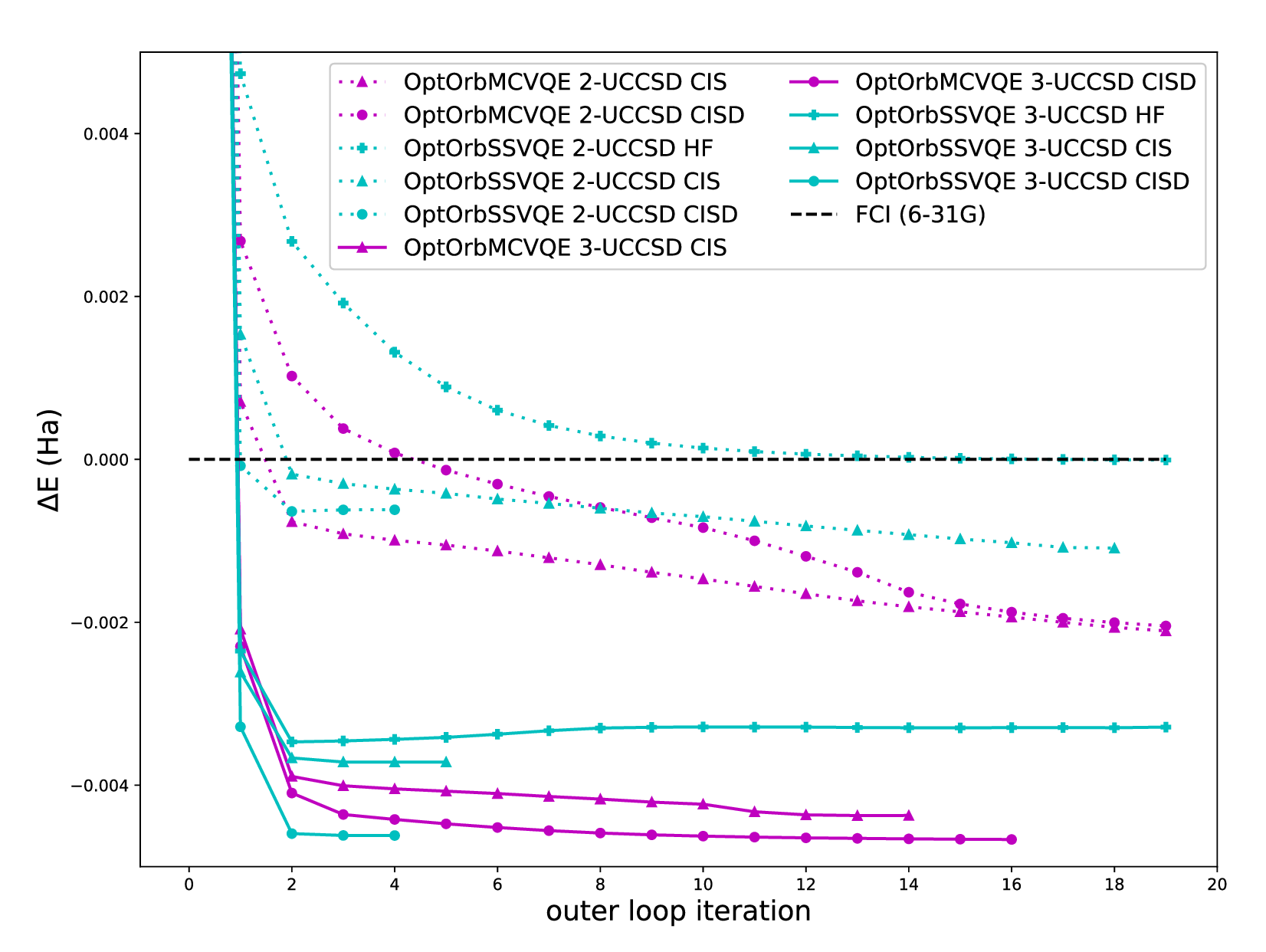}
    \caption{Convergence of orbital optimization methods for $\ch{H_4}$
    using 8 optimized spin-orbitals (taken from the cc-pVQZ basis) as a
    function of the outer loop iteration. $\Delta E$ is the difference
    between the average energy and that of FCI in the 6-31G basis (16
    spin-orbitals).}
    \label{fig:H4_results}
\end{figure}

We now present the results for the first three energy levels of \ch{H4}, a toy system comprised of four
hydrogen atoms arranged in a square with a nearest-neighbor distance of
1.23~\AA. The starting basis set is cc-pVQZ (240 spin-orbitals) and an active space of 8 optimized spin-orbitals
is used. Both 2 and 3-UCCSD are tested as ansatzes and both CIS and CISD are
tested as initializations. SSVQE is additionally tested using the Hartree-Fock initialization. The results are given in
Fig.~\ref{fig:H4_results}.

We can see that for this system, orbital optimization can be used to
achieve a more accurate average energy than the 6-31G basis (16 spin-orbitals), despite the
fact that it is utilizing half the number of spin-orbitals. Convergence
approaching FCI cc-pVDZ (40 spin-orbitals) accuracy was not observed in our testing. Between
the three different algorithmic choices considered (the eigensolver, the
initialization, and the ansatz), increasing the ansatz expressiveness from
2-UCCSD to 3-UCCSD had the most significant effect on the converged
accuracy. Changing the initialization from CIS to CISD offered a clear
improvement when used with the 3-UCCSD ansatz, however the same is not
true for 2-UCCSD. With 2-UCCSD, OptOrbSSVQE using CISD converges quickly
to a local minimum, whereas OptOrbSSVQE using CIS converges (albeit
comparatively slowly) to a more accurate average energy. The final
converged values for OptOrbMCVQE are similar between CIS and CISD when
using 2-UCCSD. Note also that for instances using the same ansatz and
initialization, using MCVQE as the eigensolver typically offers an
improvement over SSVQE. The one exception to this is using CISD with
3-UCCSD, where the difference between these two converged values is
negligible. 

\subsection{LiH}
\label{sec: LiH}

\begin{figure}[htb]
    \centering
    \includegraphics[width=0.95\linewidth]{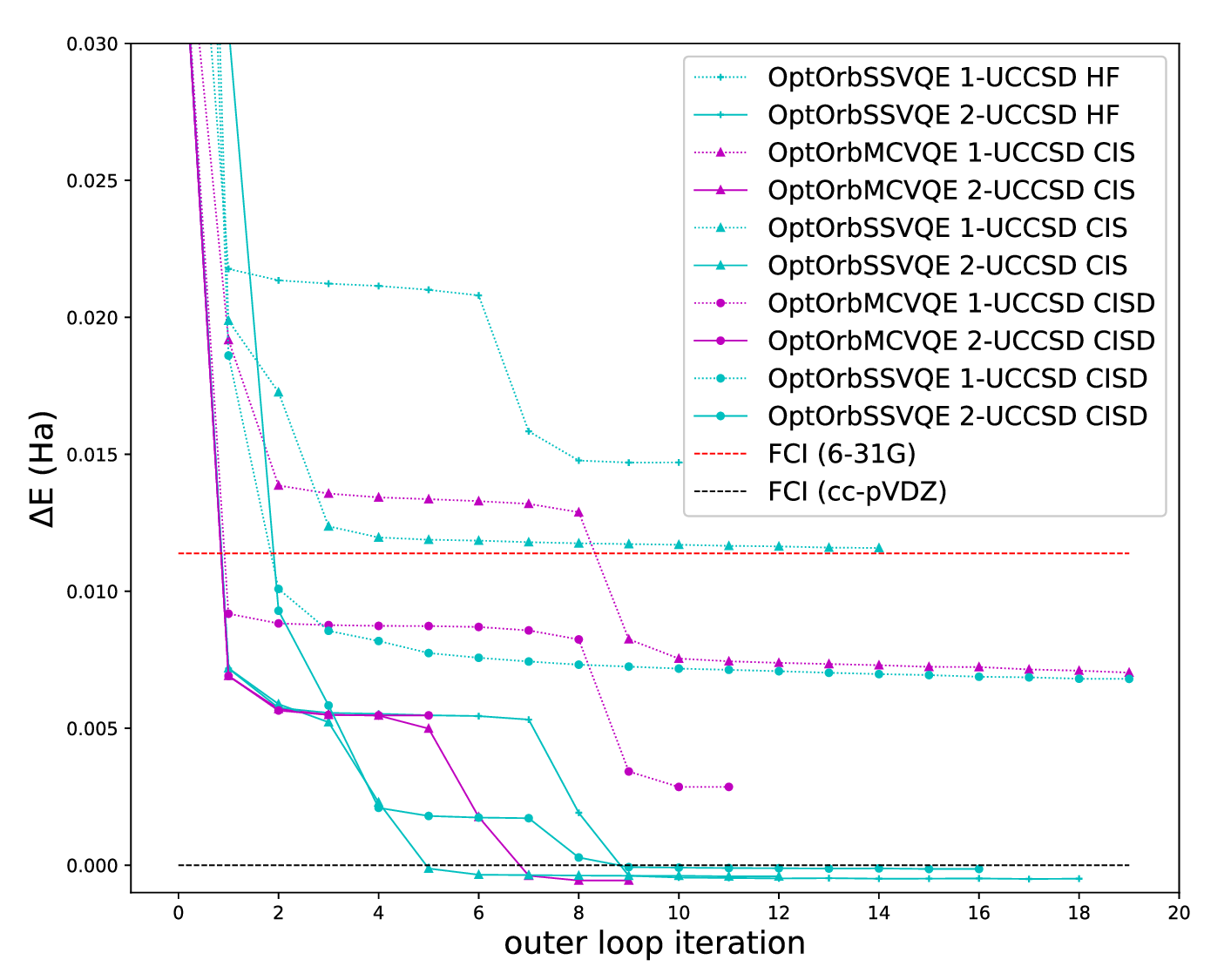}
    \caption{Convergence of orbital optimization methods for $\ch{LiH}$
    using 12 optimized spin-orbitals as a function of the outer loop
    iteration. $\Delta E$ is the difference between the average energy and
    that of FCI in the cc-pVDZ basis (36 spin-orbitals).}
    \label{fig:LiH_results}
\end{figure}

We now present the results for the first two energy levels of \ch{LiH} at
the near-equilibrium interatomic distance of 1.595~\AA. The starting basis
set is cc-pVTZ (88 spin-orbitals) and an active space of 12 optimized spin-orbitals is used. Both 1 and 2-UCCSD are
used to assess the effect of ansatz expressiveness. CIS and CISD
initializations are tested for MCVQE whereas SSVQE additionally tests the Hartree-Fock initialization. The results are shown in
Fig.~\ref{fig:LiH_results}.

The most notable feature of this plot is that orbital-optimized solvers
can achieve more accurate results than FCI using much larger basis sets.
For example, most tests run for this system outperform FCI 6-31G (22
spin-orbitals) while using only 12 spin-orbitals. Depending on the choice
of solver, ansatz, and initialization, some instances also outperform FCI
cc-pVDZ (36 spin-orbitals). The second notable
feature is that ansatz expressiveness (typically) has a greater influence
on the final accuracy than the choice of initialization. The 1 and 2-UCCSD
tests almost form two cleanly separated accuracy tiers, except for
OptOrbMCVQE using 1-UCCSD with a CISD initialization, which achieves a
higher accuracy than MCVQE using CISD with 2-UCCSD. The choice of initialization has a greater impact when the less expressive 1-UCCSD ansatz is used and has little impact when the ansatz is sufficiently expressive to approximate the solution states well. The
third notable feature is that OptOrbMCVQE typically outperforms
OptOrbSSVQE when using the same ansatz and initialization, with the one exception to this being when CISD and 2-UCCSD
are used. This effect is most noticeable when the less expressive 1-UCCSD
ansatz is used. 

\subsection{\ch{BeH2}}
\label{BeH2}

\begin{figure}[htp]
    \centering
    \includegraphics[width=\linewidth]{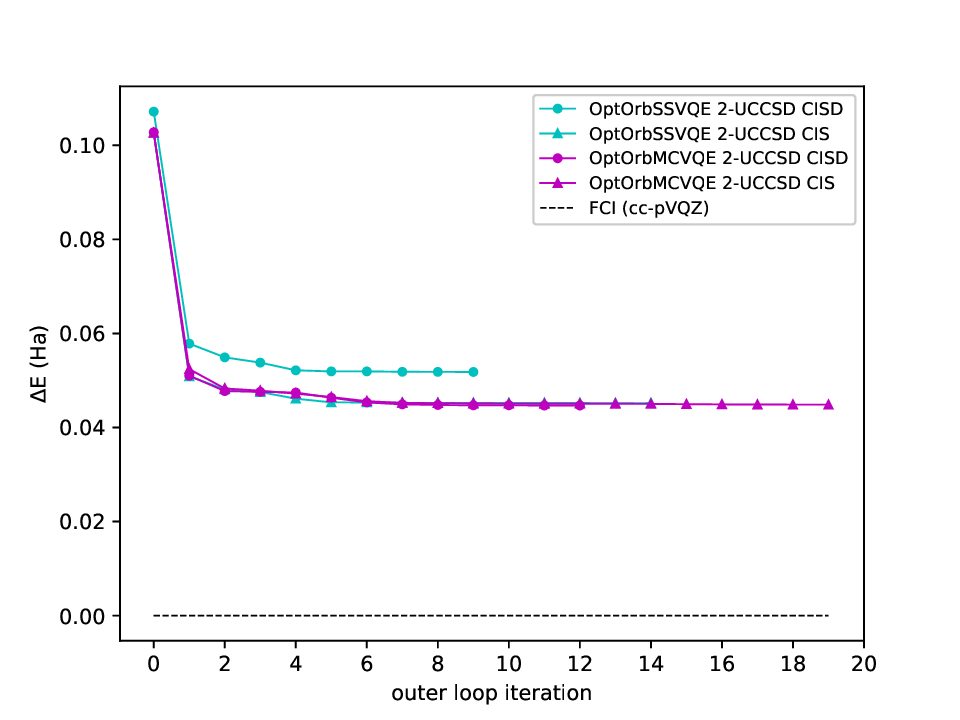}
    \caption{Convergence of orbital optimization methods for $\ch{BeH2}$
    using 12 optimized spin-orbitals as a function of the outer loop
    iteration. $\Delta E$ is the difference between the average energy and
    that of FCI in the cc-pVTZ basis.}
    \label{fig:BeH2_results}
\end{figure}

We now present the results for the first two energy levels of \ch{BeH2}
with a linear geometry at the near-equilibrium Be-H distance of 1.3264~\AA
. We find the full system with 14 spin-orbitals and 6 electrons to be
intractable for our computational budget, so we freeze two electrons in
the Hartree-Fock orbitals with the lowest energy and compare the active
space energy against that of FCI using the same frozen core approximation.
Because we do not wish for the quality of the frozen core approximation
across different basis sets to influence the comparison against FCI
values, here we will only compare the orbital optimized results starting
with the cc-pVQZ basis with an active space of 12 spin-orbitals against
FCI in the cc-pVQZ basis using an active space of 228 spin-orbitals.
Because of the wide disparity in active space size, we do not expect the
orbital optimized tests to approach chemical accuracy compared to FCI,
however we may still gain some insight as to what portion of the full
basis set energy is attainable using a small active space and what kind of
improvement orbital optimization offers over a naive approach which
chooses a fixed active space based on Hartree-Fock orbital energies. These
results are given in Fig.~\ref{fig:BeH2_results}.

\subsection{\ch{H4} Noisy Binding Curve}\label{sec: H4 binding curve}

\begin{figure*}[h!]
    \centering
    \begin{subfigure}{0.499\linewidth}
        \centering
        \includegraphics[width=\linewidth]{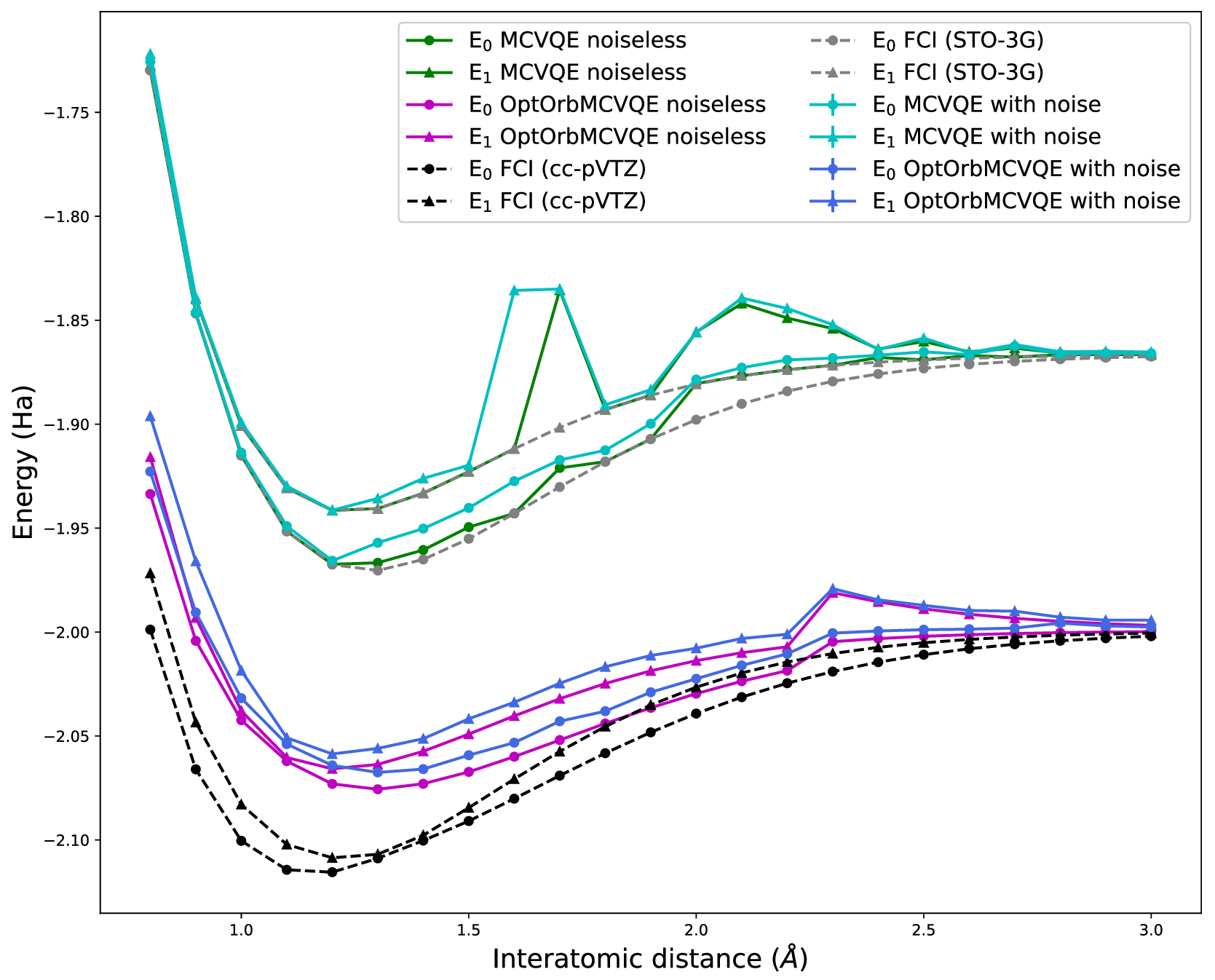}
        \caption{Absolute energies}
        \label{fig:H4_OptOrbMCVQE_binding_curve}
    \end{subfigure}
    \begin{subfigure}{0.49\linewidth}
        \centering
        \includegraphics[width=\linewidth]{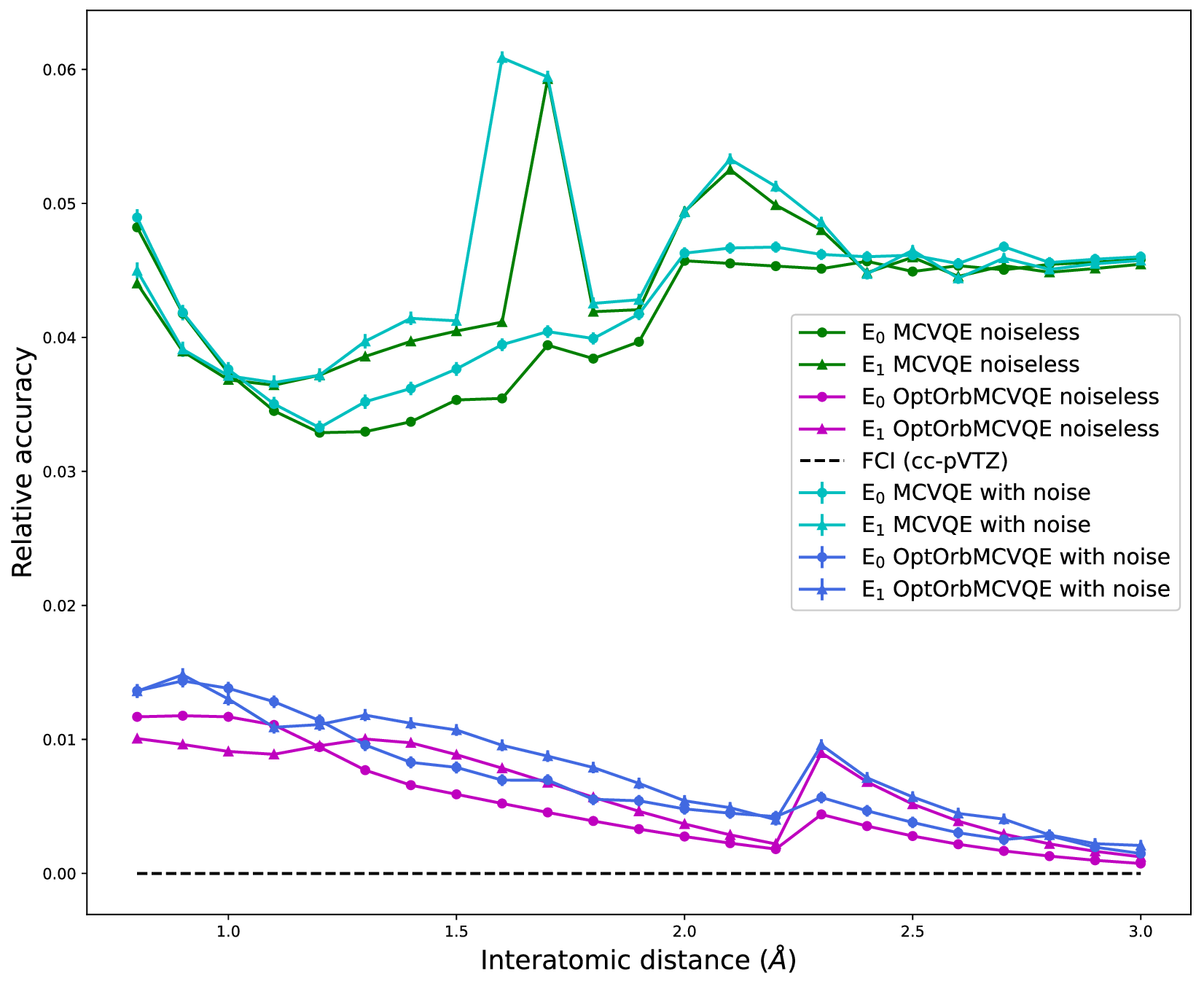}
        \caption{Relative energy accuracies}
        \label{fig:H4_OptOrbMCVQE_binding_curve_accuracies}
    \end{subfigure}
    \caption{The potential energy surface of the first two energy levels of \ch{H4} corresponding to uniformly stretching nearest-neighbor interatomic distances. a) Absolute energies for OptOrbMCVQE (cc-pVTZ, 8 spin-orbitals) and MCVQE (STO-3G, 8 spin-orbitals) along with the FCI values in the STO-3G and cc-pVTZ basis sets for comparison. b) Relative accuracies of $\mathrm{E}_0$ and $\mathrm{E}_1$ for the various methods compared to their exact FCI values in the full cc-pVTZ basis (112 spin-orbitals). Error bars are included but are small.}
    \label{fig:H4_binding_curve}
\end{figure*}

In this section we use OptOrbMCVQE to compute the potential energy surface resulting from uniformly stretching the nearest-neighbor interatomic distance of the \ch{H4} toy model. We find that this potential energy surface is difficult for state-averaged solvers such as MCVQE to accurately describe in the STO-3G basis, even when starting from CISD states. This is especially true for stretched geometries, where MCVQE often converges to local minima. Thus, this may serve as an interesting test bench for orbital-optimized state-averaged eigensolvers such as OptOrbMCVQE. We study both MCVQE (STO-3G with 8 spin-orbitals) and OptOrbMCVQE (cc-pVTZ with 8 spin-orbitals) in the noiseless case as well as in the case where noise arises from statistical sampling. These results are compared against the exact FCI values in the STO-3G and cc-pVTZ basis sets for comparison. The methodology is largely the same as in previous sections with some notable exceptions. The first is that we now use the COBYLA~\cite{powellDirectSearchAlgorithms1998} optimizer for the eigensolver subroutine as we find that it is more robust to noise than L-BFGS-B. Furthermore, we use PySCF's FCI implementation to generate exact comparison values. As opposed to previous sections where we studied various initializations and ansatz expressiveness, here we restrict ourselves to CISD initializations with the 3-UCCSD ansatz. As we will present the individual energy levels instead of the state-averaged energy, this simplifies presentation of the data. OptOrbMCVQE uses cc-pVTZ as the staring basis and uses an active space of 8 spin-orbitals. MCVQE uses the 8 spin-orbital STO-3G basis. For the tests which incorporate statistical sampling noise, we use $10^6$ shots per observable evaluation using Qiskit's approximate treatment of sampling noise, which assumes a Gaussian distribution for observable measurements and returns the variance $\sigma$. We estimate the error associated with these measurements for a number of shots $n$ as $\sqrt{\frac{\sigma}{n}}$. We find that the variance returned by Qiskit is typically on the order of $10^{-1}$. Fig.~\ref{fig:H4_OptOrbMCVQE_binding_curve} shows the absolute energies for the binding curve for the first two energy levels of \ch{H4} for OptOrbMCVQE and MCVQE. The STO-3G (8 spin-orbitals) and cc-pVTZ (112 spin-orbitals) are included for comparison. Fig.~\ref{fig:H4_OptOrbMCVQE_binding_curve_accuracies}  Note that OptOrbMCVQE uses cc-pVTZ as its starting basis but uses an active space of 8 spin-orbitals. Thus, these (along with the other tests in this work) represent an aggressive reduction in active space size. We can see that while OptOrbMCVQE offers a significant increase in the asbolute energies over MCVQE in the fixed STO-3G basis, its relative accuracy compared to the full cc-pVTZ value is still on the order of $10^{-2}$ for many geometries. This demonstrates that although orbital optimization offers a compact basis set representation, basis sets beyond the minimal basis will still clearly be required in order to approach chemical accuracy in the infinite basis set limit. Interestingly, we also note that the potential energy surface of the orbital-optimized tests are much more smooth in comparison to the fixed-basis MCVQE tests. MCVQE in the STO-3G basis clearly struggles to reliably converge to the exact values, particularly for interatomic distances of 1.6~\AA~ and larger. The jump in energies around 2.0~\AA~ for OptOrbMCVQE similarly indicates that the method is likely converging to local minima for stretched bond distances, although this effect is qualitatively less severe than in the fixed basis case. We also note that as one would expect, the effect of noise arising from statistical sampling is to degrade the energy accuracies by an amount on the milli-Hartree level. Encouragingly, the majority of the additional accuracy attained through orbital optimization persists despite the presence of this noise.

\section{Discussion and Conclusions}
\label{sec: Discussion and Conclusions}

In this paper we have proposed an orbital optimization scheme which uses a
state-averaged approach to compute excited states of electronic structure
Hamiltonians. We have shown that this method can achieve more accurate
results than FCI using much larger fixed basis sets. We have also
investigated the effects of the choice of quantum eigensolver, ansatz
expressiveness, and state initialization. While exceptions to these trends
can be found in our results, we can make the following general observations:

\begin{itemize}
    \item Increasing the ansatz expressiveness offers the most significant
    effect among these factors.

    \item MCVQE often offers an improvement in accuracy over SSVQE for
    lower ansatz expressiveness. When higher expressiveness is used, the
    difference is often less significant.

    \item CIS initializations often offer an improvement over Hartree-Fock
    initializations, however the advantage of using CISD over CIS is
    unclear. There are several instances of CISD initialized tests
    achieving a lower accuracy than their CIS (and even Hartree-Fock)
    counterparts.
\end{itemize}

The first of these is not surprising. The ansatz expressiveness is what
primarily determines the variational flexibility of the quantum
eigensolver at each outer loop iteration. The second observation can be
explained by noting that for a given initialization and ansatz, the
solution space of SSVQE is more restricted than MCVQE. The solution space
of SSVQE with unequal weights consists of the low-lying eigenvectors
themselves, whereas the solution space of MCVQE consists of the subspace
spanned by the low-lying eigenvectors. MCVQE utilizes a post-processing
step involving a low-dimensional diagonalization problem in this subspace.
This additional variational flexibility may ease the convergence process
and allow it to partially compensate for an insufficiently expressive
ansatz. The third point, while less easily explained than the other two,
can be conjectured about.  While the initialization does have an effect on
ansatz expressiveness as it determines which excitation operators in the
UCCSD ansatz act non-trivially, is not as variationally flexible as the
parameterized ansatz itself is. Furthermore, this state is computed in the
initial basis set guess, which is usually low-quality compared to the
optimized basis set. Thus there is no guarantee that CISD computed in this
initial basis will continue to be advantageous over CIS for successive
basis set rotations. This is less likely to be the case when compared to
the Hartree-Fock initial state, which consists of only one Slater
determinant and remains the same in the Jordan-Wigner qubit encoding for
all basis sets.

One compelling and well-motivated extension of this work would be to take
a state-specific orbital optimization approach rather than a
state-averaged one. State-specific orbital optimization (as the name
implies) optimizes a different basis set for each excited state
individually rather than optimizing one basis set for an ensemble of
excited states by minimizing its average energy. State-specific orbital
optimization has been developed in the context of classical orbital
optimization algorithms~\cite{yalouzOrthogonallyConstrainedOrbital2023},
however this particular method relies on a full CI expansion of the wavefunction at every
outer loop iteration. In the quantum computing setting, such an explicit
wavefunction expansion (as opposed to the expectation value of observables
used here) would involve exponentially costly tomography and classical
storage. These CI wavefunction expansions are used to compute the overlap
of two different excited states in two different basis sets and uses them
to enforce their orthogonality. In the quantum computing setting, one
would have to develop a method which can compute these overlaps without
access to CI expansions or which does not require overlaps at all. This is
an interesting problem and will be a direction of future investigation.

\begin{acknowledgement}
    The work of JB and JL are supported by the US National Science
    Foundation under awards CHE-2037263 and DMS-2309378. YL is supported
    in part by the National Natural Science Foundation of China (12271109)
    and Shanghai Pilot Program for Basic Research - Fudan University
    21TQ1400100 (22TQ017).
\end{acknowledgement}

\bibliography{reference}

\appendix

\section{Excited States Initializations and Ansatz Expressiveness}

Here we test the effects of various initialization choices and levels of
ansatz expressiveness on the convergence of MCVQE and SSVQE in a fixed minimal basis. These tests
serve to illustrate our motivation for our particular choices in the
orbital optimized tests in Section~\ref{sec: Numerical Results} of the
main text. By ``initialization'' we mean the choice of non-parameterized
subcircuit prepended to the parameterized ansatz. The ansatz parameters
themselves are initialized to zero as this corresponds to the identity
subcircuit. Thus, this allows us to explore various chemically motivated
initializations. MCVQE is tested with configuration interaction singles
(CIS) and configuration interaction singles and doubles (CISD) state
initializations. SSVQE is tested with CIS and CISD as well as an ``excited
Hartree-Fock'' initialization used in a previous study by the
authors.~\cite{biermanQuantumOrbitalMinimization2022}. This initialization
applies single-particle fermionic excitations to the Hartree-Fock state
and chooses the resulting Slater determinants with the lowest energy to
initialize the circuit. Such states are orthogonal and can thus be used
with both MCVQE and SSVQE. The ansatz expressiveness is varied by varying
the number of times the base UCCSD circuit pattern is repeated, where we
denote the circuit consisting of $n$ UCCSD repetitions as $n$-UCCSD. L-BFGS-B is the optimizer used for these tests.

Table~\ref{tab:H4_accuracies_table} shows the final average energy
accuracy for the first three states of \ch{H4} at a nearest neighbor
distance of 1.23~{\AA} for various choices of eigensolver, state
initialization, and UCCSD expressiveness. We can see that Hartree-Fock and
CIS initializations fail to produce an accuracy greater than $10^{-2}$ Ha
for any eigensolver or level of ansatz expressiveness. Furthermore,
increasing the ansatz expressiveness offers no meaningful improvement for
these initializations. On the other hand, the CISD initialization offers
the ability to achieve greater than chemical accuracy. With 2-UCCSD, both
eigensolvers fall just short of chemical accuracy, but increasing the
ansatz to 3 and 4-UCCSD offers further improvements. Thus, we can see that
there is motivation for developing circuits which correspond to CISD
states.

\begin{table*}[htb]
    \centering
    \begin{tabular}{ccccccc}
        \toprule
        Eigensolver & Initialization & \multicolumn{4}{c}{UCCSD Repetitions} \\
        \cmidrule{3-6}
        & & 1-rep & 2-rep & 3-rep & 4-rep \\
        \toprule
        MCVQE & CIS & $4.25\times10^{-2}$ & $3.85\times10^{-2}$ & $3.70\times10^{-2}$ & $3.70\times10^{-2}$ \\
        MCVQE & CISD & $5.16\times10^{-3}$ & $2.16\times10^{-3}$ & $2.53\times10^{-4}$ & $3.73\times10^{-8}$ \\
        SSVQE & HF & $5.78\times10^{-2}$ & $4.67\times10^{-2}$ & $3.70\times10^{-2}$ & $3.70\times10^{-2}$ \\
        SSVQE & CIS & $4.26\times10^{-2}$ & $3.81\times10^{-2}$ & $3.70\times10^{-2}$ & $3.70\times10^{-2}$ \\
        SSVQE & CISD & $5.41\times10^{-3}$ & $2.45\times10^{-3}$ & $1.56\times10^{-4}$ & $3.67\times10^{-8}$ \\
        \bottomrule
    \end{tabular}
    \caption{The final accuracy of the average energy for \ch{H4} for
    given choices of eigensolver, initialization, and UCCSD ansatz
    expressiveness.}
    \label{tab:H4_accuracies_table}
\end{table*}

\begin{figure}[htb]
    \centering
    \includegraphics[width=\linewidth]{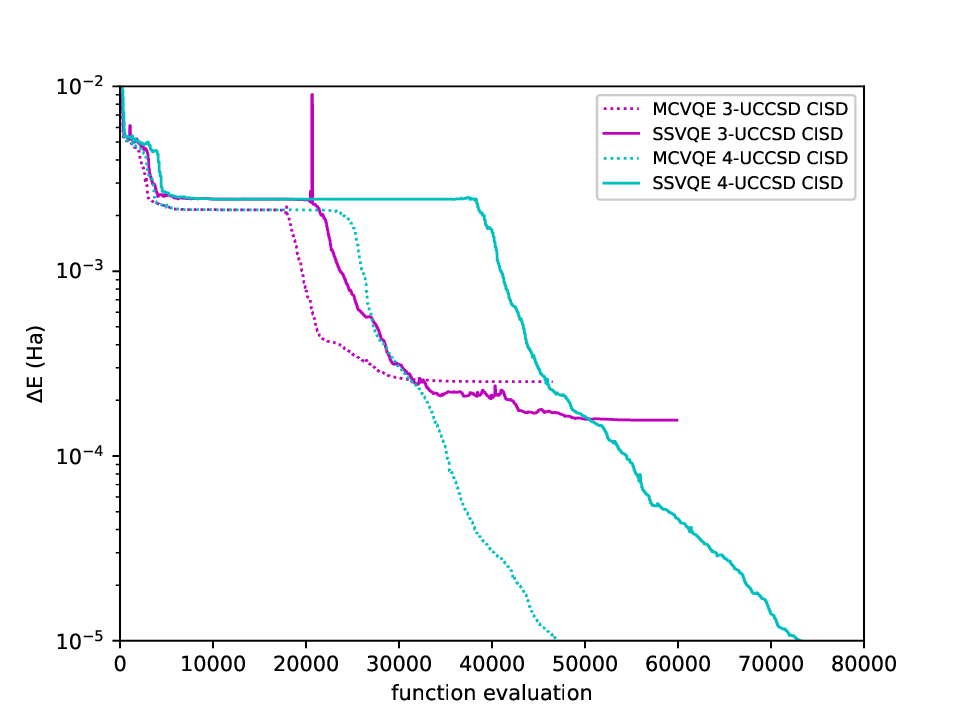}
    \caption{Convergence of the state-averaged energy accuracy ($\Delta E$)}
    \label{fig:H4_MCVQE_SSVQE_comparison}
\end{figure}

We now compare the speed of convergence between MCVQE and SSVQE for the
four test instances in Fig.~\ref{fig:H4_MCVQE_SSVQE_comparison} which were
able to surpass chemical accuracy.
Fig.~\ref{fig:H4_MCVQE_SSVQE_comparison} plots the state-averaged energy
accuracy as a function of the number of objective function evaluations. We
can see that for all four instances, the state-averaged energy plateaus
for many iterations before escaping and converging to (or closer to) the
global minimum. This is consistent with previous studies which include
SSVQE by the authors~\cite{biermanQuantumOrbitalMinimization2022}.
Notably, MCVQE is less prone to this issue.

\section{CIS State Preparation}
\label{CIS state preparation}

Here we give an example of how configuration interaction singles (CIS)
states can be prepared as a quantum circuit on a quantum computer using
Givens rotations. It was proven that Givens rotations form a universal set
of gates for chemically-motivated
statevectors~\cite{arrazolaUniversalQuantumCircuits2022}. The authors
accomplish this constructively by giving a procedure for preparing an
arbitrary state using Givens rotations controlled on the states of
multiple qubits. They comment that for particular classes of states the
resources involved may be reduced by controlling the rotation only on
certain qubit subsets. What remains to be done is to work out the details
of how to apply this idea to specific classes of CI statevectors (CIS,
CISD, CISDT, ect...) in a way that is as gate-efficient as possible. Here
we give an example of how both dense and sparse CIS statevectors can be mapped to quantum
circuits using Givens rotations.

We briefly note that the CIS state preparation circuit outlined in the MCVQE proposal paper~\cite{parrishQuantumComputationElectronic2019} assumes a particular encoding where the reference state from which electrons are being excited is encoded as the "all-zero" state $\ket{00...0}$ where the qubit registers encode the occupation number of orbitals unoccupied in the reference state, but not those occupied in the reference state. Thus, the singly-excited wavefunction components contain no information about the particular Hartree-Fock occupied orbital from which the electron was excited. Here we seek a CIS state preparation circuit in the Jordan-Wigner encoding where the reference state is the Hartree-Fock state and the occupation number of orbitals occupied in this state are included for all wavefunction components. Thus, each singly-excited wavefunction component does contain information about the occupied Hartree-Fock orbital from which the electron was excited.

The matrix representation of a Givens rotation involving qubits $n$ and
$m$ with angle $\theta$ is given by~\cite{arrazolaUniversalQuantumCircuits2022}:
\begin{equation}
    G_{nm}(\theta) = \begin{pmatrix}
        1 & 0 & 0 & 0\\
        0 & \cos{\theta} & -\sin{\theta} & 0\\
        0 & \sin{\theta} & \cos{\theta} & 0\\
        0 & 0 & 0 & 1
    \end{pmatrix}
\end{equation}
where the basis ordering is: $\ket{0}_{m}\ket{0}_{n}$,
$\ket{0}_{m}\ket{1}_{n}$, $\ket{1}_{m}\ket{0}_{n}$,
$\ket{1}_{m}\ket{1}_{n}$. or notational convenience, we will often omit
the subscript $n$ and $m$ labels on qubit registers. We can also make use
of Givens rotations controlled on the state of a target qubit $t$, which
we denote by $C_{t}G_{nm}(\theta)$. This gate can be represented as:

\begin{align}
    C_{t}G_{nm}(\theta) &= \ket{0}\bra{0}_{t}\otimes\hat{I}_{nm}\\
    &+ \ket{1}\bra{1}_{t}\otimes G_{nm}(\theta).
\end{align}
We also note that we adopt the convention of Qiskit where in the
Jordan-Wigner encoding, the qubits are ordered according to spin and
Hartree-Fock energy. Orbitals with the same spin are ordered from right to
left in ascending Hartree-Fock energy. Thus, the relevant action of a
Givens rotation is:
\begin{equation}
    G_{nm}(\theta)\ket{01} = \cos(\theta)\ket{01} + \sin(\theta)\ket{10}
\end{equation}
We do not have to consider the action of Givens rotations on the state
$\ket{10}$ as we are only interested in exciting particles to orbitals of
higher energies from lower ones. The circuit notation for the
single-excitation Givens rotation is given by~\cite{arrazolaUniversalQuantumCircuits2022}:

\begin{equation}\label{eq: Givens_gate}
    \Qcircuit @C=1em @R=1em {
        \lstick{\ket{0}} & \gate{G} &  \qw \\
        \lstick{\ket{1}} & \gate{G} \qwx & \qw \\
    }
\end{equation}
We want to construct a circuit from Givens rotations which produces the
state:
\begin{equation}\label{eq: CI expansion}
    \ket{\CIS} = C_{\HF}\ket{\HF} + \sum_{p \in \mathcal{O}_{\HF},
	q \in \mathcal{U}_{\HF}} C_{p}^{q}\ket{\phi_{q} \leftarrow \phi_{p}}
\end{equation}
where $\ket{\phi_{q} \leftarrow \phi_{p}}$ means the computational basis
state produced by exciting an electron from orbital $\phi_p$ to orbital
$\phi_q$ from the Hartree-Fock ground state. $\mathcal{O}_{\HF}$ and
$\mathcal{U}_{\HF}$ denote the set of orbitals occupied and unoccupied in
the Hartree-Fock state, respectively. We can solve for the coefficients
$C_{p}^{q}$ classically then set them equal to the parameterized
coefficients of the wavefunction expansion produced by a circuit comprised
of Givens rotations. This produces a set of equations which can be solved
to find the Givens angles which produce the circuit that prepares
arbitrary CIS states.

\subsection{Example 1: 3 particles, 6 spin-orbitals}

We now give an example for the particular case where we want to generate
the CIS wavefunction with 6 spin-orbitals and 3 particles, where all
possible single-particle excitations are considered. The circuit for
accomplishing this is given by Eq.~\ref{eq: 6o3e CIS circuit}:
\newpage
\begin{widetext}
    \begin{equation}\label{eq: 6o3e CIS circuit}
        \Qcircuit @C=1em @R=1em {
            \lstick{a \ket{1}} & \qw & \qw & \qw & \targ & \ctrl{3} & \ctrl{2} & \ctrl{1} & \targ & \ctrl{3} & \ctrl{2} & \ctrl{1} & \targ & \ctrl{4} & \targ & \targ & \targ & \qw \\
            \lstick{q_5 \ket{0}} & \qw & \qw  & \gate{G} & \qw & \qw & \qw & \gate{G} & \qw & \qw & \qw & \gate{G} & \qw & \qw & \qw & \qw & \qw & \qw \\
            \lstick{q_4 \ket{0}} & \qw & \gate{G} & \gate{G} \qwx & \qw & \qw & \gate{G} & \gate{G} \qwx & \qw & \qw & \gate{G} & \gate{G} \qwx & \qw & \qw & \qw & \qw & \qw & \qw \\
            \lstick{q_3 \ket{0}} & \gate{G} & \gate{G} \qwx & \qw & \qw & \gate{G} & \gate{G} \qwx & \qw & \qw & \gate{G} & \gate{G} \qwx & \qw & \qw & \qw & \qw & \qw & \qw & \qw \\
            \lstick{q_2 \ket{1}} & \gate{G} \qwx & \qw & \qw & \ctrlo{-4} & \qw \qwx \qw & \qw & \qw & \qw & \qw \qwx & \qw & \qw & \qw & \gate{P} & \qw & \qw & \ctrl{-4} & \qw \\
            \lstick{q_1 \ket{1}} & \qw & \qw & \qw & \qw & \gate{G} \qwx & \qw & \qw & \ctrlo{-5} & \qw \qwx & \qw & \qw & \qw & \qw & \qw & \ctrl{-5} & \qw & \qw \\
            \lstick{q_0 \ket{1}} & \qw & \qw & \qw & \qw & \qw & \qw & \qw & \qw & \gate{G} \qwx & \qw & \qw & \ctrlo{-6} & \qw & \ctrl{-6} & \qw & \qw & \qw \\
        }
        \end{equation}
\end{widetext}where the register labelled as $a$ is an ancilla qubit and those labelled
as $q_i$ are data qubits used to store the CIS state. CNOT gates with an
open dot instead of the typical filled dot denote a CNOT gate where the
NOT operation is controlled on the target qubit being in the state
$\ket{0}$ instead of $\ket{1}$. Although it is not explicitly given in the
circuit due to space constraints, each Givens rotation has its own
parameter. The controlled phase gate $P$ (implicitly $P(\lambda))$ is
given in matrix form by:
\begin{equation*}
    P(\lambda) =  \begin{pmatrix}
        1 & 0 & 0 & 0\\
        0 & 1 & 0 & 0\\
        0 & 0 & 1 & 0\\
        0 & 0 & 0 & e^{i\lambda}
    \end{pmatrix}
\end{equation*}
where the columns and rows are ordered as: $\ket{00}, \ket{01}, \ket{10},
\ket{11}$. We will see later that we only need $\lambda$ to be $0$ or
$\pi$. $\lambda = 0$ corresponds to a 2-qubit identity gate, in which case
we could omit this gate entirely, whereas $\lambda = \pi$ corresponds to a
controlled-Z gate. We denote this gate as $P$ in order to keep full
generality. The purpose of the final sequence of CNOT gates is to disentangle the ancilla
qubit from the data qubits, putting the final state in the form
$\ket{\CIS}\otimes\ket{0}$. The final state of the data
qubits is given by Eq.~\ref{eq: 6o3e CIS state}:

\begin{equation}\label{eq: 6o3e CIS state}
    \begin{split}
        \ket{\CIS} = & e^{i\lambda}\cos{\theta^3_2}\cos{\theta^3_1}\cos{\theta^3_0}\ket{000111} \\
        & + \cos{\theta^3_2}\cos{\theta^3_1}\sin{\theta^3_0}\cos{\theta^4_0}\ket{001110} \\
        & + \cos{\theta^3_2}\cos{\theta^3_1}\sin{\theta^3_0}\sin{\theta^4_0}\cos{\theta^5_0}\ket{010110} \\
        & + \cos{\theta^3_2}\cos{\theta^3_1}\sin{\theta^3_0}\sin{\theta^4_0}\sin{\theta^5_0}\ket{100110} \\
        & \\
        & + \cos{\theta^3_2}\sin{\theta^3_1}\cos{\theta^4_1}\ket{001101} \\
        & + \cos{\theta^3_2}\sin{\theta^3_1}\sin{\theta^4_1}\cos{\theta^5_1}\ket{010101} \\
        & + \cos{\theta^3_2}\sin{\theta^3_1}\sin{\theta^4_1}\sin{\theta^5_1}\ket{100101} \\
        & \\
        & + \sin{\theta^3_2}\cos{\theta^4_2}\ket{001011} \\
        & + \sin{\theta^3_2}\sin{\theta^4_2}\cos{\theta^5_2}\ket{010011} \\
        & + \sin{\theta^3_2}\sin{\theta^4_2}\sin{\theta^5_2}\ket{100011}. \\
    \end{split}
\end{equation}
We denote the angle which first adds the component $\ket{\phi_q \leftarrow
\phi_p}$ to the overall wavefunction as $\theta^q_p$.  By setting these
coefficients equal to those of the CI wavefunction expansion given in
Eq.~\ref{eq: CI expansion}, we arrive at the following set of equations in Eq.~\ref{eq: 6o3e equations}:

\begin{equation}\label{eq: 6o3e equations}
	\begin{split}
	    & e^{i\lambda}\cos{\theta^3_2}\cos{\theta^3_1}\cos{\theta^3_0} = C_{\HF} \\
	    & \cos{\theta^3_2}\cos{\theta^3_1}\sin{\theta^3_0}\cos{\theta^4_0} = C^3_0 \\
	    & \cos{\theta^3_2}\cos{\theta^3_1}\sin{\theta^3_0}\sin{\theta^4_0}\cos{\theta^5_0} = C^4_0 \\
	    & \cos{\theta^3_2}\cos{\theta^3_1}\sin{\theta^3_0}\sin{\theta^4_0}\sin{\theta^5_0} = C^5_0 \\
	    & \\
	    & \cos{\theta^3_2}\sin{\theta^3_1}\cos{\theta^4_1} = C^3_1 \\
	    & \cos{\theta^3_2}\sin{\theta^3_1}\sin{\theta^4_1}\cos{\theta^5_1} = C^4_1 \\
	    & \cos{\theta^3_2}\sin{\theta^3_1}\sin{\theta^4_1}\sin{\theta^5_1} = C^5_1 \\
	    & \\
	    & \sin{\theta^3_2}\cos{\theta^4_2} = C^3_2 \\
	    & \sin{\theta^3_2}\sin{\theta^4_2}\cos{\theta^5_2} = C^4_2 \\
	    & \sin{\theta^3_2}\sin{\theta^4_2}\sin{\theta^5_2} = C^5_2. \\
	\end{split}
\end{equation}
The recursive structure of this circuit allows us to solve for all of these
parameters analytically in a recursive way. We can partition these 10
equations into 3 blocks of 3 equations and one block with one equation
according to the occupied Hartree-Fock orbital from which the excitations
are generated. We start with $p = 2$ and solve for $\theta^5_2$, $\theta^4_2$, and
$\theta^3_2$ in that order. This has the solution:

\begin{equation}
	\begin{split}
	    & \theta^5_2 = \arctan{\left(\frac{C^5_2}{C^4_2}\right)} \\
	    & \theta^4_2 = \arctan{\left(\frac{1}{\cos{\theta^5_2}}\frac{C^4_2}{C^3_2}\right)} \\
	    & \theta^3_2 = \arcsin{\left(\frac{C^3_2}{\cos{\theta^4_2}}\right)}. \\
	\end{split}
\end{equation}
The equations corresponding to $p = 1$ are the same in structure to those
of $p = 2$, except that the left hand side is multiplied by a constant
factor of $\cos{\theta^3_2}$, a quantity that we solved for in the $p = 2$
equations. We define $\alpha_2 = \cos{\theta^3_2}$ and divide both sides
of these equations by $\alpha_2$. We arrive at a second set of solutions:
\begin{equation}
	\begin{split}
	    & \theta^5_1 = \arctan{\left(\frac{C^5_1}{C^4_1}\right)} \\
	    & \theta^4_1 = \arctan{\left(\frac{1}{\cos(\theta^5_1)}\frac{C^4_1}{C^3_1}\right)} \\
	    & \theta^3_1 = \arcsin{\left(\frac{1}{\alpha_2}\frac{C^3_1}{\cos(\theta^4_1)}\right)}. \\
	\end{split}
\end{equation}

The $ p = 0$ block of equations also has the same form, but the left side
is multiplied by a factor of $\alpha_1\alpha_2 =
\cos{\theta^3_1}\cos{\theta^3_2}$. We divide by sides of each equation in
this block by $\alpha_1\alpha_2$ and arrive at the solution for this third
block:
\begin{equation}
	\begin{split}
	    & \theta^5_0 = \arctan{\left(\frac{C^5_0}{C^4_0}\right)} \\
	    & \theta^4_0 = \arctan{\left(\frac{1}{\cos{\theta^5_0}}\frac{C^4_0}{C^3_0}\right)} \\
	    & \theta^3_0 = \arcsin{\left(\frac{1}{\alpha_1\alpha_2}\frac{C^3_0}{\cos{\theta^4_0}}\right)}. \\
	\end{split}
\end{equation}
This leaves only the parameter $\lambda$ for which to solve. The magnitude of
$C_{\HF}$ will match that of $\alpha_2\alpha_1\alpha_0$ due to the
normalization condition, but the two may differ by a factor of either $+1$
or $-1$. The parameter $\lambda$ will determine this phase. If the phase
of the two quantities match, then $\lambda = 0$ and the phase gate can be
omitted entirely. If the two differ by a phase of $-1$, then $\lambda =
\pi$.

\subsection{Example 2: Sparse 2 Particles, 6 Spin-Orbitals}

The previous example dealt with the particular case of 3 particles and 6
spin-orbitals where every single-particle excitation from any occupied
Hartree-Fock orbital is possible. We now give an example for a different
number of particles and spin-orbitals for the case where the CIS
wavefunction is sparse and some of the coefficients are zero. This
demonstrates that we can also generate approximate CIS wavefunctions at
lower circuit depth in a straightforward, systematic way by omitting
certain excitations if their CI coefficients are below a specified
threshold.

Here we suppose that we are preparing a CIS state of a system with 2
particles and 6 spin-orbitals, where $\phi_1$ can only be excited to
$\{\phi_2$, $\phi_5\}$ and $\phi_0$ can only be excited to $\phi_4$. The
circuit for doing so is given by Eq.~\ref{eq: 6o2e sparse CIS circ}:

\begin{widetext}
\begin{equation}\label{eq: 6o2e sparse CIS circ}
    \Qcircuit @C=1em @R=1em {
        \lstick{a \ket{1}} & \qw & \qw & \targ & \ctrl{2} & \targ & \ctrl{5} & \targ & \targ & \qw \\
        \lstick{q_5 \ket{0}} & \qw & \gate{G} & \qw & \qw & \qw & \qw & \qw & \qw & \qw \\
        \lstick{q_4 \ket{0}} & \qw & \qw \qwx & \qw & \gate{G} & \qw & \qw & \qw & \qw & \qw \\
        \lstick{q_3 \ket{0}} & \qw & \qw \qwx & \qw & \qw \qwx & \qw & \qw & \qw & \qw & \qw \\
        \lstick{q_2 \ket{0}} & \gate{G} & \gate{G} \qwx & \qw & \qw \qwx & \qw & \qw & \qw & \qw & \qw \\
        \lstick{q_1 \ket{1}} & \gate{G} \qwx & \qw & \ctrlo{-5} & \qw \qwx & \qw & \gate{P} & \qw & \ctrl{-5} & \qw \\
        \lstick{q_0 \ket{1}} & \qw & \qw & \qw & \gate{G} \qwx & \ctrlo{-6} & \qw & \ctrl{-6} & \qw & \qw \\
    }
\end{equation}
\end{widetext}

This results in the data qubits being put in the state:

\begin{equation}
    \begin{split}
        \ket{\CIS} = & e^{i\lambda}\cos{\theta^2_1}\cos{\theta^4_0}\ket{000011} \\
        + & \cos{\theta^2_1}\sin{\theta^4_0}\ket{010010} \\
        + & \sin{\theta^2_1}\sin{\theta^5_1}\ket{100001} \\
        + & \sin{\theta^2_1}\cos{\theta^5_1}\ket{000101} \\
    \end{split}
\end{equation}
This leads to the set of equations:
\begin{equation}
    \begin{split}
        e^{i\lambda}\cos{\theta^2_1}\cos{\theta^4_0} &= C_{\HF} \\
        & \\
        \cos{\theta^2_1}\sin{\theta^4_0} &= C^4_0 \\
        & \\
        \sin{\theta^2_1}\cos{\theta^5_1} &= C^2_1 \\
        \sin{\theta^2_1}\sin{\theta^5_1} &= C^5_1 \\
    \end{split}
\end{equation}
We solve for $\theta^5_1$, $\theta^2_1$, $\theta^4_0$, and $\lambda$
recursively in that order. The solution is given by:
\begin{equation}
    \begin{split}
        & \theta^5_1 = \arctan{\left(\frac{C^5_1}{C^2_1}\right)} \\
        & \theta^2_1 = \arcsin{\left(\frac{C^2_1}{\cos{\theta^5_1}}\right)} \\
        & \theta^4_0 = \arcsin{\left(\frac{C^4_0}{\alpha_1}\right)} \\
        & e^{i\lambda} = \frac{C_{\HF}}{\alpha_1\alpha_0} \\
    \end{split}
\end{equation}

\newpage
\subsection{General Procedure}

Based on the particular examples given, we can observe a general procedure
for any number of particles and spin-orbitals. We first partition the
spin-orbitals into two sets $\mathcal{O}_{\HF}$ and $\mathcal{U}_{\HF}$,
the set of spin-orbitals occupied and unoccupied in the Hartree-Fock
reference state, respectively. For each $\phi_p \in \mathcal{O}_{\HF}$, we
generate an ordered set $L_p$ of orbitals $\phi_q \in \mathcal{U}_{\HF}$ for which
the CI amplitude $C^q_p$ is not zero or is not below a desired truncation
threshold. These orbitals are ordered in ascending Hartree-Fock energy.
For every spin-orbital in each $L_p$, we map the orbital indices $q$ to
new indices $n_p^q$. This is simply so that we may write down a general
analytical expression for the gate sequence which reflects the fact we may
not want or need the full, dense CI wavefunction. $n_p^q$ is the index of
the list $L_p$ which was mapped from the original index of the
spin-orbital $\phi_q$. \emph{e.g.} the original set of unoccupied orbitals
from which a particular occupied orbital may be given by $\{\phi_3,
\phi_6, \phi_8\}$, but we map this ordered set to the list indices
$\{0,1,2\}$. 

The general sequence of gates is given by Eq.~\ref{eq: CIS general state}.
\newpage
\begin{widetext}
\begin{equation}\label{eq: CIS general state}
    \begin{split}
        \ket{\CIS} = &\left[\prod_{\phi_p \in \mathcal{O}_{\HF}} \CNOT(a, p)\right]C_aP(\lambda)_0\\
        & \times \prod_{\phi_p \in \mathcal{O}_{\HF}\setminus \{\phi_0\}}^{q_{max}-1}\left[X_p\CNOT(a,p)X_p\prod_{\phi_q \in L_p}C_aG_{n_p^q, n_p^q+1}\right]\\
        & \times X_0\CNOT(a,0)X_0\prod_{\phi_q \in L_0}^{q_{max}-1} G_{n_0^q, n_0^q+1}(\theta^q_0)\ket{\HF}\ket{1}_a
    \end{split}
\end{equation}
\end{widetext}
The rightmost terms denote the fact that for the set of excitations from
the first occupied spin-orbital, we do not have to apply the Givens
rotations conditioned on the state of the ancilla qubit. Without loss of
generality, we may take $\phi_0$ to be the orbital which has the longest
list $L_p$ of possible excited orbitals. This will reduce the circuit
depth as compiling controlled Givens rotations into a sequence of 1 and
2-qubit basis gates will in general be more expensive than regular Givens
rotations. After this, we apply a NOT gate to the ancilla qubit
conditioned on the state of the qubit from which we just generated
excitations being $\ket{0}$. This marks all data qubit wavefunction
components that are not the Hartree-Fock component so that future Givens
rotations will not apply excitations to these components. We then repeat
this with Givens rotations controlled on the state of the ancilla for all
the other Hartree-Fock occupied orbitals. If there is an orbital for which
there are no possible excitations, we simply skip it. We then apply a
phase gate $P(\lambda)$ to any of the Hartree-Fock occupied orbitals
conditioned on the state of the ancilla. This applies the relative phase
$e^{i\lambda}$ to the Hartree-Fock component of the wavefunction. Finally,
for each of the Hartree-Fock occupied orbitals, we apply a NOT gate
controlled on the state of the ancilla. This disentangles the data qubits
from the ancilla qubit so that the final result is a product state of
these registers.

Finally, we give a general procedure for mapping the CI coefficients $C^q_p$ to the Givens rotation
angles $\theta^q_p$. In order to do this,
we temporarily re-index the CI coefficient indices in the same way that we
did for the general circuit expression. For each $L_p$, we map the orbital
indices $q \rightarrow n^q_p$. Here, $C^{n^q_p}_p$ is the re-indexed CI
coefficient mapped from $C^q_p$ for list $L_p$. The sequence of steps for
this procedure can be given by:

\begin{enumerate}
    \item For each $\phi_p \in \mathcal{O}_{\HF}$ (In the corresponding
    order applied in the circuit):
    \begin{enumerate}[labelindent=0pt,wide]
        \item If length($L_p$) = 1:
        \begin{equation*}
            \theta^{n^q_p}_p = \arcsin{\left(\frac{C^{n^q_p}_p}{\prod_{p^{\prime}<p}\alpha_{p^{\prime}}}\right)}
        \end{equation*}
        \item If length($L_p$) = 2:
        \begin{equation*}
            \begin{split}
                & \theta^{n^q_p=1}_p = \arctan{\left(\frac{C^{n^q_p=1}_p}{C^{n^q_p=0}_p}\right)} \\
                & \theta^{n^q_p=0}_p = \arcsin{\left(\frac{1}{\prod_{p^{\prime}<p}\alpha_{p^{\prime}}}\frac{C^{n^q_p=0}}{\cos{\left(\theta_p^{n^q_p=0}\right)}}\right)}
            \end{split}
        \end{equation*}
        \item If length($L_p$) > 2:
        {\allowdisplaybreaks
        \begin{eqnarray*}
            \theta^{n^{q_{max}}_p}_p & = & \arctan{\left(\frac{C^{n^{q_{max}}_p}_p}{C^{n^{q_{max}}_p-1}_p}\right)} \\
            \theta^{n^{q_{max}}_p-1}_p & = & \arctan{\left(\frac{1}{\cos{\left(\theta^{n^{q_{max}}_p}_p\right)}}\frac{C^{n^{q_{max}}_p-1}_p}{C^{n^{q_{max}}_p-2}_p}\right)} \\
            & \vdots & \\
            \theta^{n^q_p}_p  & = & \arctan{\left(\frac{1}{\cos{\left(\theta^{n^{q}_p + 1}_p\right)}}\frac{C^{n^{q}_p}_p}{C^{n^{q}_p-1}_p}\right)} \\\
            & \vdots & \\
            \theta^{n^q_p=0}_p & = & \arcsin{\left(\frac{1}{\prod_{p^{\prime}<p}\alpha_{p^{\prime}}}\frac{C^{n^q_p=0}}{\cos{\left(\theta_p^{n^q_p=0}\right)}}\right)}
        \end{eqnarray*}
        }
    \end{enumerate}

    \item Solve for $\lambda$:
    \begin{equation*}
        \lambda=
        \begin{cases}
            0, & \text{if}\ \frac{C_{\HF}}{\prod_p \alpha_p}=1 \\
            \pi, & \text{if}\ \frac{C_{\HF}}{\prod_p \alpha_p}=-1
        \end{cases}
    \end{equation*}
\end{enumerate}
Here, for the sake of notational convenience we define $\prod_{p^{\prime}
< 0} \alpha_{p^\prime} = 1$.

\end{document}